\documentclass[reprint,twocolumn,amsmath,amssymb,aps,prx,floatfix]{revtex4-2}

\usepackage{xcolor}
\usepackage{graphicx}%
\usepackage{dcolumn}%

\usepackage{amsmath}
\usepackage{amssymb}
\usepackage{graphicx}
\usepackage{booktabs}
\usepackage{braket}
\usepackage{chngcntr}
\definecolor{NewBlue}{rgb}{0, 0, 0.41}
\definecolor{NewRed}{rgb}{0.6, 0.07, 0.07}
\usepackage{libertine}
\usepackage[colorlinks,
    linkcolor=NewBlue,
    citecolor=NewBlue,
    urlcolor=NewBlue]{hyperref}

\begin{document}
\newcommand{\C}{$^{13}\mathrm{C}\ $}
\newcommand{\TTwoStar}{$T_2^*\ $}
\title{Two-qubit cryo-CMOS control of a spin register in diamond}

\author{H. B. van Ommen$^{1,2}$}
\thanks{These authors contributed equally to this work.}
\author{N. E. Fakkel$^{1,3}$}
\thanks{These authors contributed equally to this work.}
\author{M. van Riggelen$^{1,2}$}
\author{J. Yun$^{1,2}$}
\author{L. A. Enthoven$^{1,3}$}
\author{M. A. B. ElBadry$^{1,3}$}
\author{F. Sebastiano$^{1,3}$} %
\author{M. Babaie$^{1,4}$} %
\author{T. H. Taminiau$^{1,2}$}
 \email[Corresponding author. Email: ]{T.H.Taminiau@TUDelft.nl}

\affiliation{$^1$QuTech, Delft University of Technology, 2628 CJ Delft, The Netherlands}

\affiliation{$^2$Kavli Institute of Nanoscience Delft, Delft University of Technology, 2628 CJ Delft, The Netherlands}

\affiliation{$^3$Department of Quantum and Computer Engineering, Delft University of Technology, 2628 CD Delft, The Netherlands}

\affiliation{$^4$Department of Microelectronics, Delft University of Technology, 2628 CD Delft, The Netherlands}

\date{\today}
\begin{abstract}
Utility-scale quantum technologies will not only require a well-performing quantum platform, but also classical control hardware that is scalable, efficient, and meets the design requirements set by the quantum system of interest. One approach is the application of silicon CMOS technology, a highly scalable and mature technology, to design application-specific integrated circuits (ASIC) that are optimized for the quantum system. In this work, we present two-qubit control of electron and nuclear spins in NV centres in diamond, using a cryogenic CMOS qubit controller integrated with a diamond quantum device in a cryostat. Using Gate Set Tomography (GST) we characterize the performance of the single-qubit gates on both electron- and carbon-nuclear spins, and analyze the limiting noise processes.
We use dynamical decoupling to engineer a two-qubit gate, and estimate $F_{2q,\text{avg}}=95(2)\%$ for the electron-nuclear two-qubit gate fidelity. Our demonstration of cryo-CMOS-driven two-qubit gates is a key step towards highly-integrated quantum devices based on colour centres.
\end{abstract}

\maketitle

\section*{Introduction}
There is an ongoing shift in quantum technologies from laboratory experiments to system-level engineering efforts~\cite{abrahamDigitallyControlledSilicon2026, ransford98qubitTrappedionQuantum2026, underwood2024using}. Translating lab results to robust, practical, and deployable systems not only requires the quantum platforms themselves to be mature; the entire system control stack must be examined. The integration of quantum devices and classical control systems presents a co-design problem, in which each quantum platform presents its own unique challenges.

For many platforms, a major challenge to scaling up to large systems is posed by the control complexity and low operating temperatures. Early demonstrations rely on room-temperature control systems, outside the cryostat, with each qubit typically requiring one or more control channels. Realistically, a fault-tolerant system will require error-corrected logical qubits, supported by $\sim$$10^6$ physical qubits~\cite{babbush_securing_2026}, though specific architectures may require only $\sim$$10^4$~\cite{cain_shors_2026}. Scaling towards this number of qubits therefore also scales the engineering complexity and added heat load due to the cryostat interconnects (e.g. optical fibers, coaxial cables)~\cite{krinner2019engineering}.

Across multiple qubit platforms, one proposed solution is to place the control system in close proximity to the quantum device: inside the cryostat. This can be achieved with application-specific integrated circuits (ASIC) using conventional silicon CMOS technology, which is both highly scalable, and has been shown to work reliably at cryogenic temperatures~\cite{Patra2018}. Cryogenic CMOS (cryo-CMOS) integrated control systems can be designed in a highly specialized fashion towards the design requirements posed by the quantum system of interest. This targeted cryo-CMOS design approach enables miniaturization of a large portion of the control system, offering a significant reduction in form factor~\cite{xue2021cmos}, a lower thermal noise floor~\cite{bardin_design_2019}, and better power efficiency~\cite{Mehrpoo2019BenefitsComputers} than room-temperature approaches, and can satisfy the low-latency feedback requirement for quantum error correction~\cite{knapen_pinball_2026}. Cryo-CMOS control of superconducting qubits~\cite{underwood2024using, bardin_design_2019}, silicon quantum-dot qubits~\cite{blumoff2026digitallycontrolled,park2021fully}, and trapped ions~\cite{toth202513} has recently been demonstrated, with each quantum platform presenting its own unique design challenges and opportunities.  

An interesting class of qubit platforms is formed by solid-state defect centres, or colour centres. They are formed by a point defect in a crystal. We consider the nitrogen-vacancy (NV) centre in diamond. The NV centre features long electron-spin coherence times at room temperature, forming a promising quantum sensing platform~\cite{degen2017sensing}, and at cryogenic temperatures (below $\sim$$8$~K) they can be used as processor nodes in quantum networks~\cite{ruf_quantum_2021}, with high-fidelity quantum control and long memory lifetimes demonstrated~\cite{bartling_universal_2025,bradley_ten-qubit_2019, riggelen_high-fidelity_2026}. Due to their operating temperature, typically around 4~K, where cryostats offer Watts of cooling power, NV centres facilitate the co-integration of the qubit array and the cryo-CMOS controller at the same temperature.%

A quantum processor platform based on solid-state defects faces specific challenges towards system-level miniaturization.
First, large alternating current (AC) magnetic field driving is required to achieve fast, high-fidelity gates based on magnetic resonance.
While room-temperature electronic control integration efforts exist~\cite{li2024heterogeneous}, driving long 50-$\Omega$ transmission lines inherently results in bulky and power-hungry driver solutions.
Second, optical initialization and readout of, and generating photon-mediated entanglement between many colour-centre network nodes will also require integrated photonics to achieve system-level miniaturization~\cite{Ishihara20213DQubits}. 
With recent advances in silicon–photonic integration, CMOS can already provide a monolithic solution for photonic control~\cite{10727757}.
Thus, a promising pathway towards scalable quantum systems based on solid-state defects is to use a cryo-CMOS qubit controller, integrated with the diamond quantum sample at the same temperature stage.

\begin{figure}
    \centering
    \hspace*{-9mm}
    \includegraphics[width=1.15\columnwidth]{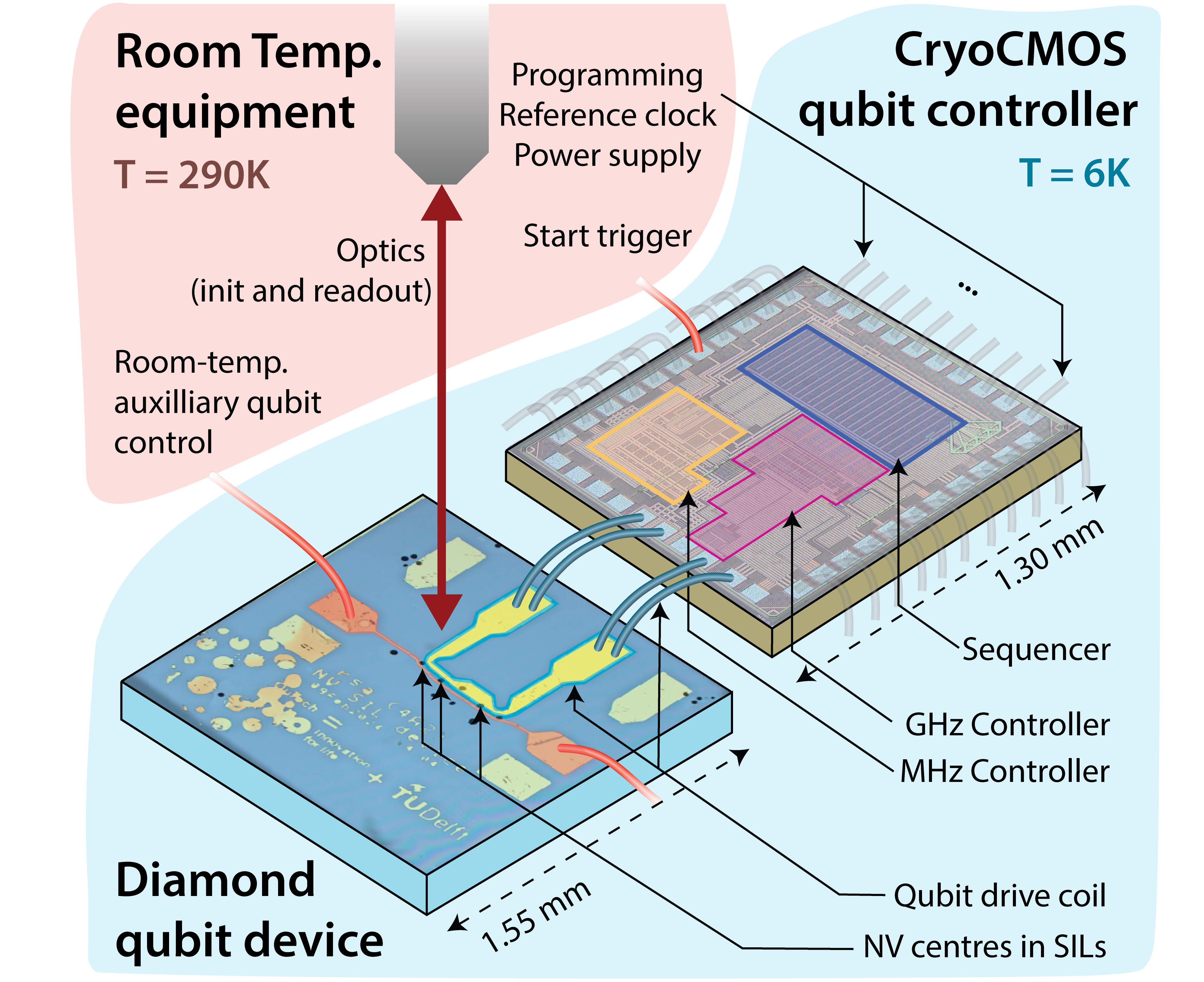}
    \caption{\emph{Integrated qubit system.} The cryo-CMOS qubit controller chip~\cite{enthovenCryoCMOSElectronNuclearSpin2026} and diamond qubit device hosting NV centres are placed together on a printed circuit board in a cryostat at 6~K. Solid Immersion Lenses are used to improve the optical collection efficiency of the NV centre emission. The CMOS chip autonomously performs a quantum circuit on the qubits. Qubit initialization and readout are performed by the room-temperature control system. Prior to circuit runtime the CMOS chip is programmed to perform the desired quantum circuit(s). A reference clock, power supply, and a starting trigger are provided by external room-temperature systems. A magnetic field of 196 mT is applied perpendicular to the diamond surface using a permanent magnet in the cryostat.}
    \label{fig:fig1}
\end{figure}

In this work, we present multi-qubit control of spin qubits associated to an NV centre in diamond, controlled by a cryo-CMOS application-specific integrated circuit (ASIC), see Fig.~\ref{fig:fig1}. The detailed electronical design and basic testing of the cryo-CMOS ASIC used have been previously presented in Enthoven \textit{et al.}~\cite{enthovenCryoCMOSElectronNuclearSpin2026}. Compared to that work, here we realize universal control over two-qubits by designing, implementing and demonstrating a two-qubit gate. Additionally, we discuss the design rationale of the chip and the interplay of the electronics design with the physics and quantum-control requirements of the NV centre, as well as with different quantum-gate designs. Finally, we consider which design choices should be prioritized in future chip iterations to improve fidelity and scale control to larger qubit spin-qubit registers.

\section{System Architecture}

\begin{figure*}[ht]
    \centering
    \includegraphics{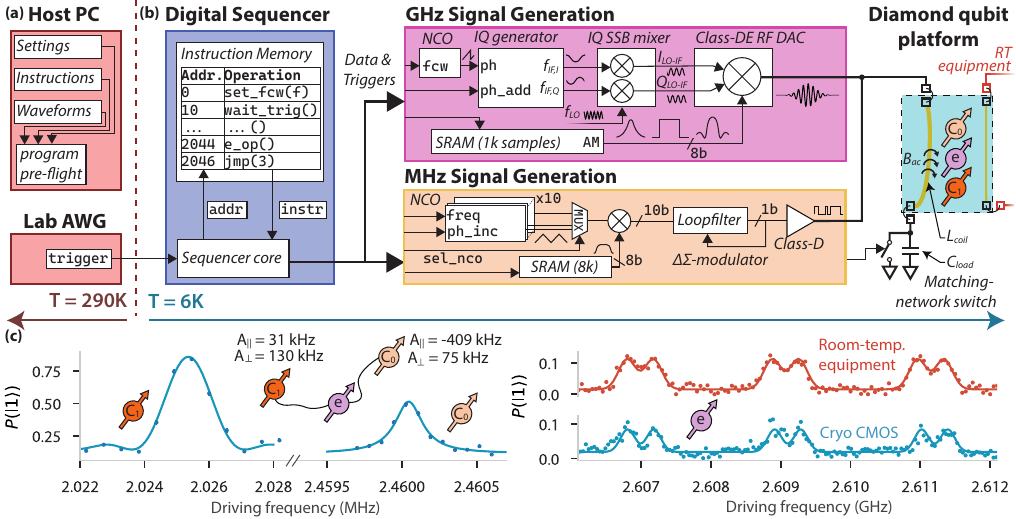}
    \caption{\emph{CMOS controller architecture.} \textbf{a)} The qubit controller is programmed prior to runtime by the host PC. This includes static configuration of the signal generators, the sequencer instruction memory, and the waveform memory. \textbf{b)} The digital sequencer follows the instructions contained in instruction memory. Most instructions lead to the writing of data to a configuration register (such as changing the GHz signal frequency) or triggering of either signal generation path to start waveform playback. The sequencer can be triggered by an external system to start executing instructions. The GHz signal generator controls the electron-spin qubit. In-phase (I) and Quadrature-phase (Q) signals are synthesized from a Numerically Controlled Oscillator (NCO) at an intermediate frequency $f_{IF}\approx20$~MHz, and up-converted to the qubit frequency $f_e\approx2.6$~GHz by single-sideband (SSB) mixing with the external reference clock at frequency $f_{LO}$. The signal is amplified by a class-DE RFDAC, controlled by 8-bit waveform samples from the waveform SRAM. The MHz signal generator controls the nuclear-spin qubits. The presence of 10 NCOs allows tracking of up to 10 spins. The triangle-wave output of one of the NCOs can be selected using the multiplexer (MUX), which is then modulated by waveform SRAM, digitized to a 1-bit signal using a loopfilter, and amplified by a class-D amplifier. A single qubit-driving coil is used. A matching-network switch is present to optimize the impedance for either MHz or GHz driving. An auxiliary driving coil for the room-temperature equipment is present for initialisation and readout. \textbf{c)} Measured frequency spectra of the qubit register using the CMOS (blue) or room-temperature (red) control system. The electron-spin qubit spectrum splits into six peaks due to the spin-1 nitrogen (not used in this work) and the strongly-coupled $c_0$ spin. inset: hyperfine coupling strengths between the qubits.}
    \label{fig:architecture}
\end{figure*}

In this section, we summarize the integrated system comprising a cryo-CMOS qubit controller chip and a diamond qubit device featuring NV centres (Fig. \ref{fig:fig1}), focusing on the capabilities for generating control pulse sequences for the NV center qubits. The detailed electronics design and testing of the chip is presented in Enthoven et al.~\cite{enthovenCryoCMOSElectronNuclearSpin2026}. 

The cryo-CMOS chip and the diamond quantum device are glued side-by-side onto a cold-head extrude and connected to a printed circuit board (PCB) by bondwires. The entire assembly is placed in a cryostat in a confocal microscope and cooled to 6~K (for details see Appendix \ref{sec:chip_integration}). The side-by-side placement provides free optical access for initialization and readout, which are orchestrated by the room-temperature control system.

The NV-centre quantum system forms a small qubit register, consisting of an electron-spin qubit $e$ which is coupled to nearby, naturally occurring \C nuclear spins $c_i$ via the hyperfine interaction~\cite{doherty_nitrogen-vacancy_2013}. The presence of two different qubit types offers unique advantages, but their distinct control frequencies and timescales also pose a challenge for the controller design. The main difference is caused by the different gyromagnetic ratios $\gamma_e=2.8$~MHz/G and $\gamma_c\approx1.07$~kHz/G~\cite{doherty_nitrogen-vacancy_2013}. At the operating magnetic field $B\approx196$~mT the qubit frequencies are $f_e\approx2.6$~GHz and $f_c\approx2$~MHz; the relatively high field is favourable for nuclear-spin control and coherence~\cite{pompili_realization_2021}. Because $\gamma_c$ is much smaller, the nuclear spins have much longer coherence times ($T_{2,c}^*\approx10$~ms, $T_{2,e}^*\approx5$~\textmu s), while control gates typically have a much longer duration compared to the electron spin ($T_{\text{gate},c}\approx100$~\textmu s, $T_{\text{gate},e}\approx100$~ns). 

The required gate fidelities for fault-tolerant quantum computation impose stringent accuracy and noise requirements on the controller's amplitude, frequency, timing, and phase, which ensure that the controller itself does not restrict the obtainable fidelities.
For example, a controller-limited fidelity above 99.99\% (i.e., infidelity $<10^{-4}$) translates to a set of control specifications \cite{enthoven2024optimizing,vandijkImpactClassicalControl2019} whose combined error and noise power must remain $\sim$40~dB below the signal power, for both nuclear and electron spin control.

The two qubit types give rise to a chip design (Fig. \ref{fig:architecture}b) with two signal generators; a GHz subsystem for controlling the single electron spin, and a MHz subsystem for controlling up to 10 nuclear spins. To improve the scalability and reduce the design complexity of the qubit device, both signal generators use the same driving-coil. The drive-coil is patterned on a diamond qubit device and directly bonded to the qubit controller. The drive-coil is designed to ensure that both signal generators can efficiently drive the coil, by incorporating it with a switchable matching network in the chip. The subsystems are controlled at runtime by the pre-programmed on-chip digital sequencer. Finally, the qubit device features a second driving line connected to a room-temperature control system~\cite{riggelen_high-fidelity_2026}, used for diagnostics, qubit initialization and readout, and enabling direct system comparisons.

\subsection{Electron spin GHz signal generation}
The GHz signal generator can be triggered by the digital sequencer to play a pulse. The frequency, phase, duration, and amplitude shape of each pulse can be controlled. We use a Single-Sideband (SSB) modulation scheme to up-convert the frequency of a Numerically Controlled Oscillator (NCO) at $f_{IF}\approx20$~MHz to the qubit frequency $f_e = f_{LO} - f_{IF}\approx2.6$~GHz. This design has multiple advantages: (1) Sideband mixing reduces LO leakage, a common issue for direct conversion transmitters~\cite{razavitx1999}, pushing leakage out-of-band and protecting the qubit; (2) The NCO enables digital control of the signal phase by the sequencer; and (3) The output frequency can be fine-tuned using the digitally controlled $f_{IF}$. This frequency flexibility is important for scalable architectures, where it can be used to compensate for magnetic field inhomogeneities across a multi-node chip~\cite{enthoven2024optimizing}.
The 15-bit frequency control word (FCW) for the NCO results in a $10$~kHz frequency resolution. 
The resulting microwave drive pulse delivered to the qubit can be expressed as
\begin{equation}
    B_{ac}(t) = \frac{\Omega_\mathrm{AM}(t)}{\gamma_e} \cdot \sin(2\pi f_e t + \varphi), 
\end{equation}
where the phase word $\varphi$ (i.e., the qubit's rotation-axis $R_\varphi$), is digitally added to the IQ generator, and the time-dependent amplitude envelope $\Omega_\mathrm{AM}(t)$ is defined by 8-bit samples stored in waveform memory. 

In typical operation, the waveform memory is pre-programmed with a library of elementary qubit operations ($R_x(\pi), R_x(\pi/2), ...$) which are chained together by the digital sequencer. The clock for the digital systems of the GHz generator, consequently also the waveform sample rate, is derived from $f_{LO}$; here $f_{LO}/8\approx312$~MHz (MSa/s) is used. The 1kS waveform memory can store up to 12 envelopes for typical gate durations. Combined with the sequencer-level phase control this enables the efficient execution of more complicated sequences, such as dynamical decoupling, which protects the electron spin coherence and which we will use in section \ref{sec:twoqubit} to demonstrate electron-nuclear two-qubit gates~\cite{taminiau_detection_2012, bartling_universal_2025, abobeih_one-second_2018}.

During GHz driving, the matching network switch is \emph{open} forming a high-Q resonator comprising the inductance of the driving line $L_{coil}$ and parasitic load capacitance $C_{load}$.
Designing the driving-coil and matching network's $L_{coil}C_{load}$-resonance frequency close to the electron transition frequency $f_e$ means that a switch-mode Class-DE driving stage can be used to reach large current amplitude with minimum power \cite{fakkel2024cryo2}. The amplitude of the pulse is controlled by implementing the Class-DE output stage as an 8-bit radio-frequency DAC (RFDAC). The RFDAC consists of 256 unit cells, which are enabled depending on the desired amplitude code ($\Omega_{\mathrm{AM}}(t)$). This architecture achieves a peak current over 80~$\mathrm{mA}_p$, which translates to an electron-spin Rabi frequency exceeding 2~MHz. In future large-scale systems, a combination of greater cryogenic cooling capacity and improved antenna–qubit integration should enable higher Rabi frequencies and correspondingly faster gate execution while reducing cryostat heating.

\subsection{Nuclear spin MHz signal generation}
The MHz signal generator also operates on a pulse level. The frequency of the signal is controlled using a 24-bit NCO operating at $f_s=4f_{\mathrm{instr}}=52$~MHz, enabling 3~Hz frequency resolution. High precision is desired due to the long nuclear-spin coherence times. The phase of the NCO can be incremented directly with a 24-bit phase word, allowing the execution of virtual-$z$ gates. The presence of 10 NCOs allows the control of up to 10 nuclear spins. The NCOs output a 10-bit triangle wave which is digitally multiplied by a fixed amplitude or by 8-bit waveform samples from the 8~KiB nuclear-spin SRAM. The sample rate is determined by $f_s/d$, where $d$ is a configurable clock division factor, so that long waveforms can be executed efficiently. 
The digital signal is quantised to 1 bit by a loopfilter, resulting in a pulse density modulated (PDM) signal. The loopfilter output is evaluated every $f_s$ clock cycle, and drives a Class-D amplifier which connects the qubit-control line to either the positive or negative supply voltage.

During MHz driving, the matching network switch is \emph{closed} forming a low impedance path to ground. Due to the shorted inductance, a high-pass current filter suppresses higher harmonics and allows for PDM amplitude control.
Using low-resistance Class-D switches enables the use of low supply voltages, reducing the power consumption, while still driving currents up to $36~\mathrm{mA}_p$.
Thus, the matching network switch provides efficient driving in both MHz and GHz operation, sharing only a single driving-coil, to support further system integration.

\subsection{Digital sequence generation}
The digital sequencer controls the settings (phase, frequency, duration, and waveform selection) of the GHz and MHz generators, and triggers them to start pulse playback at the desired time. It is programmed using a small instruction set (16 instructions); each instruction is represented by a 4-bit opcode and, for some instructions, extra data (e.g. `jmp' requires the address to jump to). The instructions are stored in a 1~KiB programmable instruction memory and are executed sequentially at runtime. During experiments, instruction-clock frequencies of either $f_{\mathrm{instr}}\sim13$~MHz or $f_{\mathrm{instr}}\sim56$~MHz were used. The combination of an auxiliary counter and a conditional jump instruction enables more complex programmes, such as long dynamical decoupling sequences, to be programmed in a memory-efficient manner, as we will highlight in the next sections. Waiting times between pulses are controlled using `nop' operations, which simply wait one $f_{\mathrm{instr}}$ cycle. Long waiting times can also be efficiently programmed by looping over a number of `nop' operations, or by repurposing the MHz signal generation path, using the $f_s$ clock from the MHz signal generator as a counter.
Prior to operation, the instruction memory is programmed by a PC. After the room-temperature control system has initialized the qubits, the sequencer is triggered to start execution of the program.

\section{Single-Qubit Control}

We previously reported single-qubit control using the cryo-CMOS controller as part of the initial chip characterisation~\cite{enthovenCryoCMOSElectronNuclearSpin2026}. In this section we first briefly recap those results, which show control of the NV-centre electron spin and two \C nuclear spins. Then, we provide an extended discussion of the electron-spin coherence under dynamical decoupling, and investigate the dominant factors limiting the fidelities of the single-qubit gates using gate set tomography (GST). Additionally, we describe 
how these operations can be efficiently programmed on the controller.

We measure the electron-spin resonance (ESR) and nuclear-magnetic resonance (NMR) frequencies of the spins by sweeping the frequency of a weak GHz (MHz) pulse using the on-board NCOs. The ESR signal shows six resonance frequencies centred at $f_e=2.6091(4)$~GHz, split by hyperfine coupling with the nitrogen nuclear spin ($n$, $A_\parallel=2.1$~MHz) and a nearby \C spin ($c_0$, $A_\parallel=-409$~kHz). For the NMR measurements, the electron spin was prepared in the $m_s=-1$ state, which separates the resonance frequencies of the nuclear spins by their hyperfine coupling: $f_c\approx\omega_l  - A_\parallel$, with Larmor frequency $\omega_l=B_z\gamma_c$. Besides the more strongly coupled $c_0$ spin with resonance frequency $f_0=2.450045(5)$~MHz we control a second spin $c_1$ with $f_1=2.02536(2)$~MHz. The parallel and perpendicular hyperfine components of both spins (Fig. \ref{fig:architecture}c) were estimated from dynamical-decoupling data \cite{taminiau_detection_2012} (see Appendix \ref{sec:hyperfine}).

The nuclear-spin Rabi frequencies lie between 1 and 2 kHz (see supp. Fig. \ref{fig:c_spin}), on-par with recent demonstrations of high-fidelity nuclear spin control \cite{van_ommen_improved_2025, riggelen_high-fidelity_2026}. The difference in maximum Rabi frequencies may be explained by the hyperfine interaction with the electron spin~\cite{doi:10.1126/science.1131871} but could partially be caused by a frequency-dependent output current of the MHz driver too.
The electron spin can be driven with a Rabi frequency of $2.31(1)$~MHz.
While this enables reasonably fast gate times, it becomes more complicated to drive the electron spin independently of the states of the strongly-coupled carbon spin $c_0$ and the nitrogen spin~\cite{vandersypen_nmr_2005, bartling_universal_2025}. Therefore we initialize the states of these spins prior to quantum circuits involving the electron spin.

\begin{figure}
    \centering
    \includegraphics[width=\columnwidth]{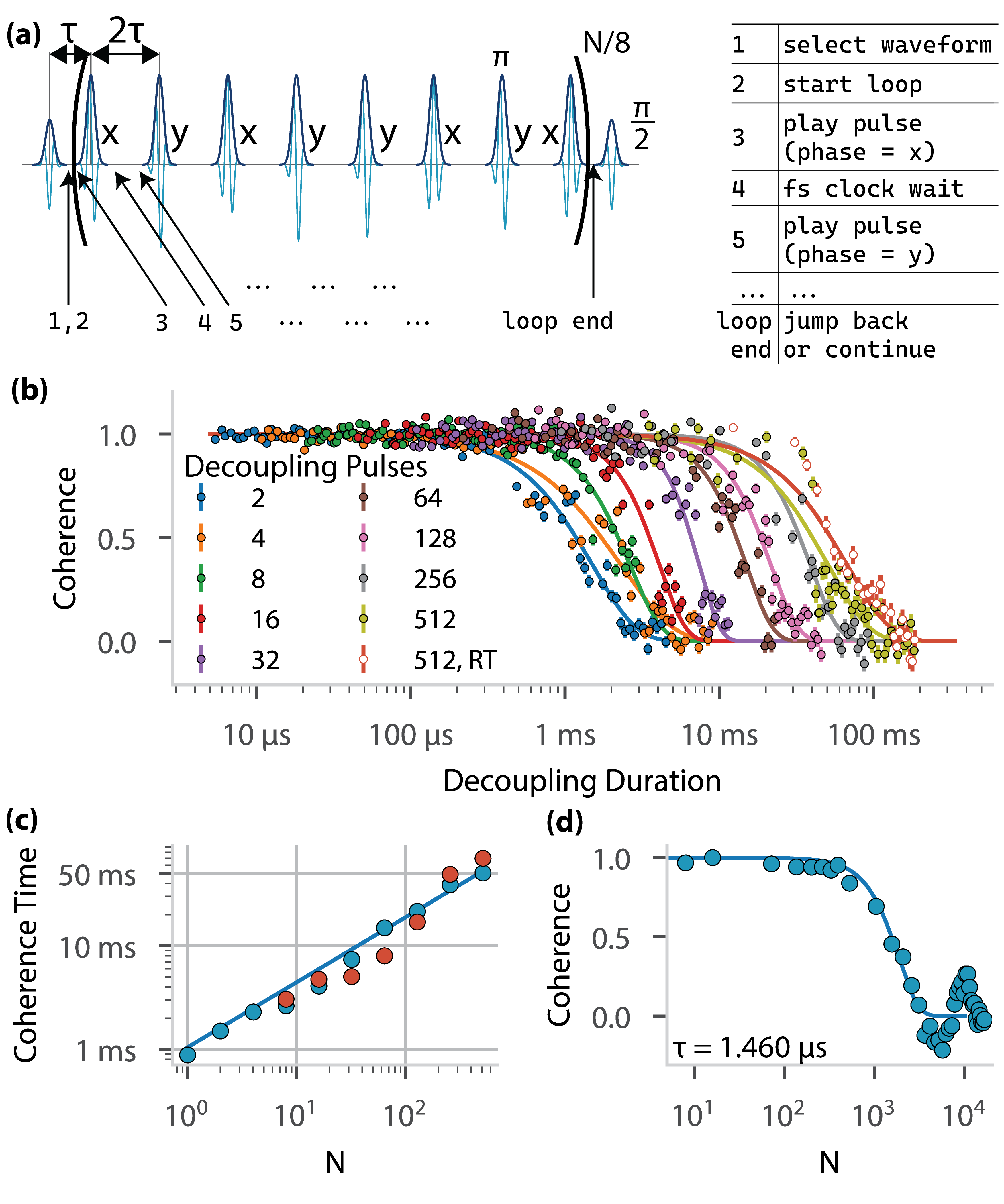}
    \caption{\emph{Electron spin decoupling.} \textbf{a)} Dynamical decoupling sequence to protect the electron-spin coherence, parametrized by $\tau$ (the time between the pulses) and $N$ (the number of pulses). Total sequence time: $T=2N\tau$. The XY8 sequence is robust to phase and amplitude errors of the $\pi$ pulses. On the right, we show the simplified steps by which the XY8 sequence is efficiently programmed. \textbf{b)} Electron-spin coherence as a function of $\tau$, for different $N$, using CMOS control. For $N=512$ a measurement using the room-temperature control system is also shown (open red circles). See Appendix \ref{sec:dd_proc} for data processing details. \textbf{c)} Scaling of the coherence time $T_2^e$ as a function of $N$. Blue (red) data is from CMOS (RT-equipment) measurements. We fit a power law $T_2^e=T_0 N^\eta$, yielding $T_0=1.1(2)$~ms and $\eta=0.62(3)$. The CMOS controller achieves similar coherence times to the RT equipment. The blue data is reproduced from our chip-characterisation work~\cite{enthovenCryoCMOSElectronNuclearSpin2026}. \textbf{d)} Coherence of the electron-spin qubit as a function of $N$ for a fixed $\tau=1.460$~\textmu s, the value used for GST experiments. Fit is $e^{-(N/N_0)^2}$ with $N_0=1.8(1)\times10^3$.}
    \label{fig:e_spin}
\end{figure}

To preserve the electron-qubit state during quantum circuits, dynamical decoupling (DD) is used, wherein the electron spin qubit is echoed periodically (Fig. \ref{fig:e_spin}a). Extending the coherence this way also enables protocols to engineer two-qubit gates between the electron and weakly-coupled nuclear spins~\cite{taminiau_detection_2012, bradley_ten-qubit_2019}. The quasi-static nature of the nuclear-spin bath makes DD highly effective, with DD coherence times ($T_2^{DD}$) that can exceed the Ramsey $T_2^*$ by more than $10^5$ times \cite{abobeih_one-second_2018, yamamoto_ten-second_2026}. 

To efficiently programme the decoupling sequence, a few measures were taken to make optimal use of the CMOS controller. The same waveform was used for the $x$ and $y$ $\pi$-pulse, so the selected waveform in the GHz generator does not have to be changed. Additionally, the `eop' instruction, which triggers the GHz generator, has a setting that allows instantaneously setting the output phase to 0, 90, 180, or 270 degrees, again saving instructions. To efficiently wait long durations between GHz pulses, we use the MHz signal generator as a counter. Finally, we use the looping functionality to repeat the XY8 sequence as required.

We perform XY8 DD (Fig. \ref{fig:e_spin}a) with up to $N=512$ pulses and show that we can preserve the electron-qubit state up to 50~ms (Fig. \ref{fig:e_spin}b,c). DD performed by the room-temperature equipment, with the cryo-CMOS powered off, shows comparable performance, demonstrating that the GHz signal generator does not affect the coherence of the electron qubit. 

\begin{figure}
    \centering
    \includegraphics[width=\columnwidth]{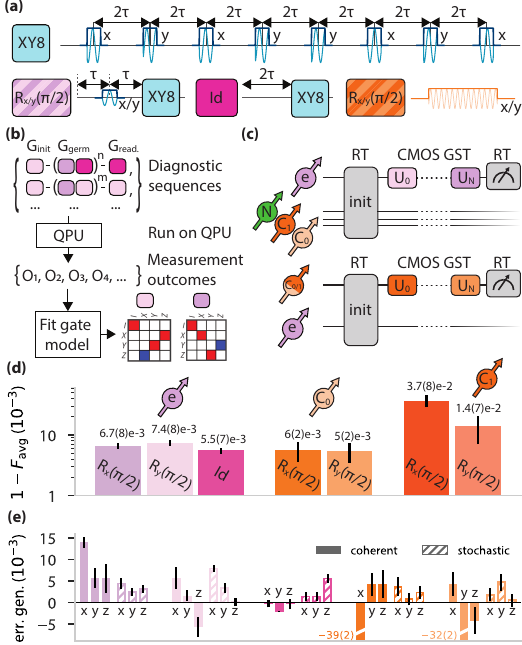}
    \caption{\emph{Gate Set Tomography.} \textbf{a)} Definition of the gate set. Electron-spin pulses are square. The identity gate consists of only an XY8 sequence, and is included to diagnose the fidelity of just the decoupling sequence. The nuclear spin gates consist of a single square pulse. \textbf{b)} GST estimates process matrices of the entire gate set (including SPAM) by fitting a model of the gate set to data obtained by running specific diagnostic sequences on the quantum processing unit (QPU). \textbf{c)} For the electron-spin GST measurements the states of $n$, $c_0$ and $c_1$ were initialized alongside $e$ by the room-temperature system. Then, the CMOS chip performed the GST sequence, after which the RT system performed qubit readout. For the carbon-spin measurements, $e$ was prepared in $m_s=-1$. \textbf{d)} Average gate infidelity for each gate, on a log scale. We considered data up to germ-length $L=16$ ($e$), $L=8$ ($c_0$), $L=2$ ($c_1$). These fidelities are also reported in our chip-characterisation work~\cite{enthovenCryoCMOSElectronNuclearSpin2026}. \textbf{e)} Error generators of the $e$ and $c_0$ gates. The coherent and stochastic error generators each have x,y, and z components. Note that, for $c_0$, the x (y) coherent error generator of the $R_x(\pi/2)$ ($R_y(\pi/2)$) gate falls off the graph. %
    }
    \label{fig:gst}
\end{figure}

The need for decoupling gives rise to a quantum circuit design where DD is constantly performed \cite{bartling_universal_2025} (Fig. \ref{fig:gst}a). Therefore we also demonstrate the spin coherence as a function of the number of $\pi$-pulses when decoupling at a fixed $\tau=1.460$~\textmu s (Fig. \ref{fig:e_spin}d). This shows that for 512 $\pi$ pulses, or 64 XY8 blocks, the electron spin is still highly coherent. Note that while a coherence decay of the form $e^{-N/N_0}$ is expected from nuclear-spin bath noise ~\cite{de_lange_universal_2010}, here the form $e^{-(N/N_0)^2}$ results in a better fit of the data, which suggests coupling to an additional noise source inside or outside the diamond.

Gate-Set Tomography~\cite{nielsen_gate_2021} is used to benchmark the single-qubit performance of the CMOS controller (Fig. \ref{fig:gst}b).
GST is a self-consistent tomography method that characterises an entire set of gates simultaneously by performing a number of diagnostic gate sequences on the quantum system. The gate sequences consist of preparation and measurement \textit{fiducials} ($G_\text{init}$, $G_\text{readout}$), which prepare and measure the qubit in a certain state, and \textit{germs} ($G_\text{germ}$), which are combinations of gates that are repeated many times (up to the total sequence length $L$) to amplify specific errors, leading to a total sequence of $G_\text{init} - (G_\text{germ})^n - G_\text{readout}$. Beyond basic metrics, such as gate fidelity, GST can provide insight into the fidelity-limiting processes by fitting the full process matrix of each gate. 

For the electron spin, the investigated gate set consists of the decoupling-protected $R_x(\pi/2)$ and $R_y(\pi/2)$ gates, which consist of an XY8 sequence followed by a x or y $\pi/2$-pulse. We also add $I_d$, the identity gate, which is only an XY8 sequence, to characterize the decoupling sequence separately (Fig. \ref{fig:gst}a).
Measurements were performed up to $L=128$, corresponding to more than 1000 physical gates, again demonstrating the flexibility of the sequencer design. However, we observed a decrease in the goodness-of-fit of the Markovian GST models when data from longer sequences are added, and therefore we consider here the data up to $L=16$ for the electron spin qubit, and $L=8$ ($L=2$) for nuclear spin $c_0$ ($c_1$) (see Appendix \ref{sec:gst_fit}).

Compared to the optimisations used for the decoupling data (Fig. \ref{fig:e_spin}), we take a slightly different approach in programming the controller. The shorter decoupling times ($\tau=1.460\mu$s) means we use `nop' instructions to wait, and we make use of the structure of the GST sequences by using the loop functionality to repeat the GST germ $G_\text{germ}$ many times (see supp. Fig. \ref{fig:cmos_programme} for an example).

The extracted electron qubit average gate fidelities exceed 99\%, the worst gate being at $99.26(8)\%$. 
The leading cause of gate infidelity can be extracted from the model by considering the error generators of each gate (Fig. \ref{fig:gst}e)~\cite{blume-kohout_taxonomy_2022}. The error generators represent the post-gate error process that maps the ideal gate to the fitted gate. We consider the coherent and stochastic components of these error generators, which represent rotations by a certain angle, and stochastic errors with a certain probability, respectively. Note that the coherent error generators only contribute in second order to $F_\text{avg}$~\cite{blume-kohout_taxonomy_2022}.
For the identity gate, the stochastic z error dominates, consistent with the picture that the XY8 sequence mitigates pulse errors well and that the electron-spin qubit mainly dephases during decoupling.
For the $\pi/2$-gates, GST reports stochastic x and y errors as the leading cause of errors, which may hint at fluctuations in the output amplitude. Notably, the reported z stochastic errors for the $\pi/2$-gates are small to negligible, which is surprising given the gate implementation which makes use of the same XY8 sequence as the $I$-gate. This could be caused by the gauge-dependence of the error generators~\cite{nielsen_gate_2021}, 
or should be seen in the context of the non-Markovian decoherence process we observed (Fig. \ref{fig:e_spin}d), which cannot be represented properly by GST.

Similarly, GST was performed on the nuclear spins. Owing to the longer coherence times the nuclear spins were not decoupled during GST, and simple square pulses were used to implement the gates. The electron-spin was prepared in $m_s=-1$. 
The average gate fidelities are above 99.4\% for $c_0$ and 96\% for $c_1$. The difference is likely related to the longer $c_1$ gate times due to the lower $\Omega_{c_1}$, and possibly the larger perpendicular hyperfine coupling $A_\perp$ of $c_1$. Considering the error generators, we see that for $c_0$, both gates have a significant under-rotation (x/y coherent error on $R_x(\pi/2)$/$R_y(\pi/2)$ gate), which may be explained by a drift in output amplitude between calibration and experiment, or a systematic calibration error which may be reduced in the future by improved calibration sequences. The fidelity is, respectively, limited by stochastic x and y noise on the $R_x(\pi/2)$ and $R_y(\pi/2)$ gates. While this may be explained by fluctuations in the output amplitude, this is unlikely since it would require fluctuations on the order of $\sim$ 5\% assuming a quasi-static amplitude for each pulse~\cite{vandijkImpactClassicalControl2019, enthoven2024optimizing}. Another possible cause are qubit-frequency fluctuations, which translate to stochastic noise along the qubit rotation axis during qubit driving.

\section{Two-qubit Control}
\label{sec:twoqubit}
\begin{figure*}
    \centering
    \includegraphics{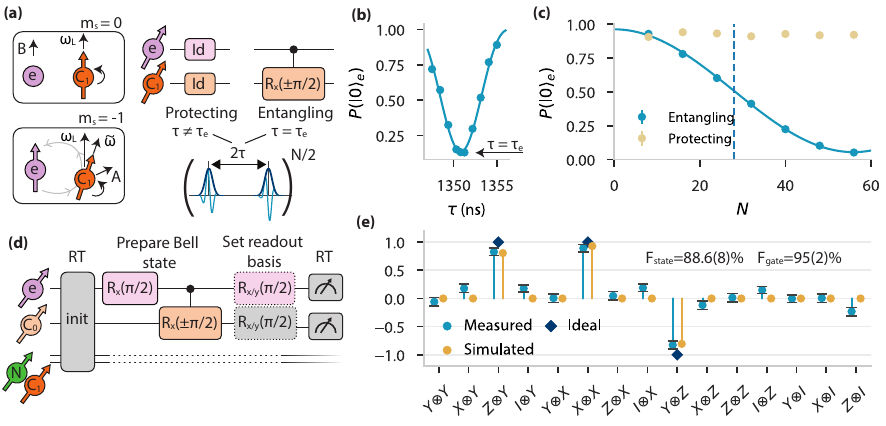}
    \caption{\textbf{Two-qubit gate}. \textbf{a)} The two-qubit hyperfine interaction between the electron spin and nuclear spin can be utilized to form a two-qubit gate by applying a DD sequences with the appropriate interpulse delay $\tau_e$. \textbf{b)} Electron-spin state after DD spectroscopy. The measurement sequence is: $R_x(\pi/2)_e - \text{DD}(\tau, N=44) - R_x(-\pi/2)_e$. Interaction between $e$ and the uninitialized $c_1$ during the DD sequence causes the characteristic dip in the DD spectrum. The centre of the dip defines the ideal $\tau=\tau_e$. \textbf{c)} To calibrate the DD gate, $N$ is swept while keeping $\tau=\tau_e$. For comparison we also show the same $N$-sweep at a different $\tau$, which simply protects the electron spin state. At $N=28$ we perform the desired gate. \textbf{d)} Quantum circuit to create and perform tomography on a Bell state. Grey blocks are performed by the room-temperature system. \textbf{e)} Bell state tomography. Expectation values are corrected for the readout infidelity of $e$ and $c_1$ (Appendix \ref{sec:readout_cor}). }
    \label{fig:bell}
\end{figure*}

Next we develop and demonstrate an electron-nuclear two-qubit gate using the CMOS controller. Such a two-qubit gate is an essential element for universal quantum control. Various gate schemes exist. For a strongly-coupled nuclear spin ($A\gg 1/T^*_{2,e}$), such as the intrinsic nitrogen spin $n$, a selective microwave pulse can be used to directly realize a $\text{C}_n\text{NOT}_e$ gate \cite{neumann_single-shot_2010, xie_9992-fidelity_2023}. Here, however, the limited maximum Rabi frequency of the qubit controller makes $n$-independent (single-qubit) control of the electron spin $e$ challenging. Instead, we focus on leveraging DD sequences to enable the control of weakly coupled spins ($A\gg 1/T^{DD}_{2,e}$) \cite{taminiau_detection_2012}. A general advantage of this method is that more nuclear spins can be controlled, and the coherence of the electron spin is protected. Here, we use a DD sequence to directly create the gate~\cite{taminiau_universal_2014}. Alternatively, the controller is capable of interleaving RF pulses in the DD sequence~\cite{bradley_ten-qubit_2019, van_ommen_improved_2025, kuna_localization_2025}, but the resulting gate fidelities were limited due to low-frequency noise (see discussion below).

First, we briefly recap how a two-qubit gate can be engineered using a DD sequence (Fig. \ref{fig:bell}a)~\cite{taminiau_detection_2012}. If the electron-spin is in the $m_s=0$ state, the hyperfine coupling is disabled and the nuclear spin precesses at a rate of $\omega_L$ along the $B$-field direction. However, with the electron-spin in $m_s=-1$, the hyperfine interaction both tilts the nuclear-spin precession axis and changes the precession rate to $\tilde{\omega}=\sqrt{(\omega_L - A_\parallel)^2 + A_\perp^2}$. In general, at high magnetic field, the tilt is small and dynamical decoupling will protect both the electron-spin and nuclear-spin state. However, when the interpulse delay $\tau$ is precisely tuned to 
\begin{align}
    \tau_e = \frac{(2k+1)\pi}{\tilde{\omega}+\omega_L},
\end{align} ($k\in\mathbf{N}$), 
the nuclear-spin rotations add up constructively, which results in an entangling gate described by the unitary \begin{align}
    U = \ket{0}\bra{0}_e \otimes R_x\left(N\frac{\phi}{2}\right)_c + \ket{1}\bra{1}_e \otimes R_x\left(-N\frac{\phi}{2}\right)_c,
\end{align} 
which is equivalent to a CNOT gate up to local operations when $N\frac{\phi}{2}=\pi/2$.

To calibrate the two-qubit gate, we first use the CMOS controller to measure the coherence of the electron spin after a decoupling sequence of $N=44$ pulses as a function of $\tau$ (Fig. \ref{fig:bell}b). The resonance feature near $\tau=\tau_e$ has a narrow linewidth which scales inversely with the magnetic field; at the field in this work it is $\sim$$4$~ns. To increase the timing accuracy of the delay $\tau$ between $\pi$-pulses, we increase the sequencer-clock frequency by a factor four compared to the single-qubit experiments, to $f_{\text{instr}}\approx56$~MHz. While this yields more consistent pulse timings, $\tau$ can still only be adjusted in coarse steps of $\sim$$18$~ns. 
We therefore perform the $\tau$ sweep by sweeping $f_{\text{instr}}$, and in all subsequent experiments set $f_{\text{instr}}$ such that $\tau$ equals $\tau_e$.
From the resonance location ($\tau_e=1.3507$~\textmu s) and depth, the hyperfine parameters $A_\parallel=31$~kHz and $A_\perp=130$~kHz can be estimated.
At the optimal $\tau_e$, we sweep the number of decoupling pulses $N$ to identify the number of pulses that results in a fully entangling gate ($N\phi/2=\pi/2$) (Fig. \ref{fig:bell}c). This occurs at $N=28$ pulses, which we implement as 3 XY8 sequences and 1 XY4 sequence.

Having calibrated the two-qubit gate, we test it by creating a Bell state between the electron- and nuclear-spin qubit (Fig. \ref{fig:bell}d). After qubit initialization using the room-temperature electronics, the Bell state is prepared by the CMOS chip. Afterwards, we perform tomography on the Bell state by measuring the 15 two-qubit Pauli correlators. The CMOS chip is responsible for setting the electron-spin readout basis, while the room-temperature equipment is used to select the nuclear-spin readout basis. We correct for the readout errors of the electron-spin and nuclear-spin qubits (see Appendix \ref{sec:readout_cor}). The resulting Bell state fidelity is $F_{state}=88.6(8)\%$, indicating two-particle entanglement. In addition, we estimate the gate fidelity of the two-qubit gate by assuming that the other performed gates are ideal (see Appendix \ref{sec:2q_fid}). This yields a lower-bound gate fidelity estimate of $F_{2q,avg}=95(2)\%$. This result demonstrates control over a two-qubit system using the cryo-CMOS chip.

\section{Discussion}

In this work, we have demonstrated two-qubit control through the integration of a cryo-CMOS qubit controller and a diamond quantum chip, as first step towards scalable integrated qubit controls. Various avenues for future research and improvements can be identified.    

Here, qubit initialization and measurement were performed by the room-temperature equipment. For a future CMOS controller to take care of these tasks a more advanced sequencer, which allows for conditional programme flow, will be required. Furthermore, extra digital input/output signals would be required to interface with the photon-detection and laser-control hardware. The CMOS chip might be used to trigger external laser-control hardware, while further integration with either an on-board laser-amplitude modulator, or perhaps complete on-chip laser systems can be investigated ~\cite{zhou_prospects_2023}. Additionally, integration with electrical readout can be explored~\cite{gulkaRoomtemperatureControlElectrical2021}.

Additional flexibility and improved control might be achieved by combining DD with with rf pulses targeting a nuclear spin (DDrf gates) \cite{bradley_ten-qubit_2019, van_ommen_improved_2025, bartling_universal_2025, riggelen_high-fidelity_2026, kuna_localization_2025}. Our qubit controller is capable of performing such DDrf gates. However, initial tests showed that the electron spin dephases rapidly while the MHz generator outputs signal. This dephasing is likely caused by the class-D design of the MHz generator, which is not filtered sufficiently, and can cause significant ac- and dc-Zeeman shifts of the electron-spin frequency. Future designs should closely consider the impact of noise emitted by the MHz generator on the electron-spin coherence.

A key challenge for the DD-based two-qubit gates presented here is the precise timing ($\lesssim$$100$~ps) of the MW pulses that is required. In this work we implemented a fine-tuned instruction clock frequency to precisely target the required pulse timing for a given nuclear-spin qubit. To control multi-qubit systems a more scalable solution is desired. 
This timing challenge can be addressed using specific digital CMOS circuits. For example, the instruction clock could be generated on chip by an all-digital phase-locked loop~\cite{8490904} to set $f_{\text{instr}}$ with fine resolution and low jitter. A digital-to-time converter~\cite{10396616} could then add a fractional delay to each gate trigger, controlling each interpulse delay with a resolution far below the clock period. In addition, to efficiently programme decoupling experiments with long waiting times between pulses, a future sequencer would benefit from a dedicated instruction to wait a specified number of instruction cycles.

Finally, large-scale control will require careful monitoring and optimisation of sample heating~\cite{fungueirino_cryo-cmos_2026}. During some of the nuclear-spin qubit experiments performed here, the electron-spin readout contrast was observed to drop (Supp. Fig. \ref{fig:e_decay_rabi}), likely caused by sample heating~\cite{fu_observation_2009}. The effect appears to be transient, as the readout contrast is restored after a $\sim$1~ms waiting time. Further reducing heating will enable higher electron-spin Rabi frequencies and will be even more important for colour-centre qubits that require dilution-refrigerator temperatures, such as group-IV colour centres in diamond~\cite{nguyen_integrated_2019, beukers_control_2025}. Avenues to reduce local heating include: (1) A silicon-photonic interposer solution places the coil closer to the defect, and the stronger coupling lowers the power dissipated by the driver\cite{psi_quantum_2025,enthovenCryoCMOSElectronNuclearSpin2026}, and (2) superconducting interconnects, such as indium ball bonds~\cite{PRXQuantum.5.030337}, would further reduce the thermal load to the sample, and lower conduction losses. Maximising Rabi frequencies while minimising dissipation thus remains an ongoing co-design challenge for the CMOS controller and the diamond chip.

In conclusion, this work has demonstrated single- and two-qubit spin control based on a cryo-CMOS controller and identified various avenues for improvement to enable higher fidelity control and more complex circuits. These future improvements include higher Rabi frequencies, more extended sequencing capabilities, control over the optical initialisation and readout, a further reduction in heat dissipation, and setting more stringent requirements for the out-of-band emission of the MHz generator to enable DDrf gates. The presented results are a step towards integrated spin-based devices for larger-scale quantum networks for quantum communication and distributed quantum computation.

\normalsize
\section*{Acknowledgements}
We acknowledge support from Zu Yao Chang from TU Delft and Nico Alberts and Tim Hiep from QuTech. We acknowledge René Vollmer from TNO for performing the diamond nanofabrication. We gratefully acknowledge support from the joint research program “Modular quantum computers” by Fujitsu Limited and Delft University of Technology, co-funded by the Netherlands Enterprise Agency under project number PPS2007. This work was supported by the Dutch National Growth Fund (NGF), as part of the Quantum Delta NL programme.  

During the preparation of this manuscript, the authors used Claude Sonnet 4.6, Sonnet 5, and Opus 5.5 to support in the refinement of text and the writing of code to produce the figures. The authors reviewed, validated and critically assessed all AI-assisted outputs and take full responsibility for the content of the manuscript.

\section*{Data Availability}
The data and code supporting the findings of this paper are available on the 4TU data server \cite{data_repo}.

\section*{Contributions}
HBvO, NF, MvR, and JY performed qubit measurements. NF, LAE, and ABEB designed the qubit controller with input from HBvO, MvR, and JY on the requirements. The manuscript was written by HBvO and NF, with input from all authors. FS, MB, and THT supervised the overall research effort.

\appendix
\makeatletter
\renewcommand{\thesubsection}{\thesection.\arabic{subsection}}
\renewcommand{\p@subsection}{}
\makeatother
\setcounter{section}{0}
\section{Experimental Methods}
\subsection{Room-temperature qubit-control system}
\label{sec:rt_sys}
Initial sample characterization, spin initialization and readout, and reference measurements were performed with a room-temperature control system, which is described in van Riggelen \textit{et al.}~\cite{riggelen_high-fidelity_2026}. The same spin-initialization and -control methods were also used. For typical experiments, after the room-temperature system has initialized the required qubits, the CMOS controller is triggered to start its sequence. The CMOS design included a trigger-out to notify the room-temperature equipment when it was done, but this trigger was faulty; the room-temperature system therefore pre-calculates the CMOS sequence duration and waits the appropriate time after triggering before starting readout. Once this waiting time has elapsed, the room-temperature system performs $Z$-readout of the required qubits, with the CMOS controller having mapped the relevant operator onto the $Z$-projection.

\subsection{Cryo-CMOS Chip Integration}
\label{sec:chip_integration}
Figure \ref{fig:cmos_setup_pictures} shows photographs of the qubit-control setup. 

A cold finger and sample holder were designed to host the PCB, the cryo-CMOS chip, and the diamond chip. The PCB has a small cut-out, so that both chips can be glued onto the cold finger while wirebonded directly to the PCB traces. The cryogenic PCB only hosts passive components for decoupling and routing of the external supply, digital, and rf signals to the chips. The assembly is mounted in a Montana Instruments Cryostation S50. The cryostat was set to a temperature of 6~K. A permanent magnet is placed inside the cold head $\sim$1~mm behind the diamond chip to produce a coarse static magnetic field. Further field alignment along the NV axis is performed using external magnets on positioner stages. A confocal microscope objective is positioned above the diamond sample.

The diamond sample contains solid immersion lenses (SILs)~\cite{hadden_strongly_2010} and NV centres (Fig.~\ref{fig:cmos_setup_pictures}c). Two coils are patterned on the diamond by gold deposition and lift-off (200-nm thickness): the qubit-drive coil $L_{\text{coil}}$ (30~$\mu$m wide), bonded to the cryo-CMOS driver output, and an auxiliary coil $L_{\text{RT}}$ (7~$\mu$m wide), connected to the RT setup. The auxiliary coil serves two purposes: (1) driving the NV cenrer from the RT rack to provide a reference measurement, and (2) feeding a Teledyne LeCroy WaveMaster oscilloscope (40\,GS/s) to monitor the line; because of capacitive coupling, the CMOS qubit-control signals could be observed in real time and used for debugging the pulse timings.

The driving-coil matching network was designed to specifically match the $L_{\text{coil}}C_{\text{load}}$ resonance frequency to the electron transition frequency $f_e$. Provided the large matching-network switch and the loading of the nuclear-spin driver, the effective on-chip capacitance was already $C_{\text{load}}=3.6$~pF. To ensure resonance around 2.5\,GHz, double bond-wires and a custom gold-metal-deposited stripline provided a sufficiently low inductance $L_{\text{coil}}=1.15$~nH. This was verified by electromagnetic (EM) simulation software.

The CMOS chip was powered by an external bench-top power supply unit (PSU) and a custom LDO board. During active driving, the CMOS drew 45~mW for the electron-spin controller from a 1.1~V supply, and 17~mW for the nuclear-spin controller from a $\pm100$~mV dual-rail supply. In idle mode, the CMOS still drew 12~mW. The clocks were provided externally by a Rohde~\&~Schwarz SMA100B for the $\sim5$~GHz clock and by a Stanford Research CG635 for the digital $\sim50$ (or 225)~MHz clock.

\subsection{Hyperfine parameter estimation}
\label{sec:hyperfine}
The coupling strengths of the nuclear spins $c_0$ and $c_1$ to the electron spin are described in terms of the parallel ($A_\parallel = A_{zz}$) and perpendicular ($A_\perp = \sqrt{A_{zx}^2 + A_{zy}^2}$) hyperfine components. These were estimated by cross-referencing dips in the DD-$\tau$ spectrum with the $A_\parallel$ values which are approximately known from the spin resonance frequencies, and matching the dip size and position by varying $A_\parallel$ and $A_\perp$ (Fig. \ref{fig:bell}b and supp. Fig. \ref{fig:dd_extended}). A nuclear spin Larmor frequency (while the electron-spin is in $m_s=0$) of $\omega_0 = 2\pi \times 2.04963$~MHz was assumed.

\subsection{Pulse calibration}
\label{sec:pulse_cal}
The calibration of $\pi$-pulses is performed by repeatedly applying the same pulse and measuring the resulting electron-spin state as a function of pulse amplitude. This process can be efficiently performed by storing pulses with different amplitudes in SRAM. We perform $\pi/2$-pulse calibration in the same fashion, but due to the nuclear-spin bath we interleave each $\pi/2$-pulse with XY8-decoupling to prevent dephasing~\cite{bartling_universal_2025}. 

\section{Data Analysis}
\subsection{DD data processing.}
\label{sec:dd_proc}
The DD data (Fig. \ref{fig:e_spin}c) was processed before fitting the coherence time curve. Considering the electron-spin coherence, the DD spectrum as a function of $\tau$ has a high density of sharp dips, due to the nuclear-spin environment (see for example Supp. Fig. \ref{fig:dd_extended}). These dips are generally avoided if the goal is to protect the electron-spin coherence, as done for example in Abobeih \textit{et al.} \cite{abobeih_one-second_2018}.
Here, for the purpose of fitting an estimated decay envelope, we excluded data according to the following rule: given the $k$-th electron coherence measurement $C_k$ at $\tau_k$, if there is a measurement $C_l>C_k+2\sigma_{C_k}+0.05$ for $\tau_l>\tau_k$, then $C_k$ is excluded. $\sigma_{C_k}$ is the measurement uncertainty in $C_k$, and the number $0.05$ is added to prevent the exclusion of too much data. The reasoning is that there exists a larger $\tau$ at which the electron maintains more coherence, so we're likely to have sampled one of the DD dips at $\tau_k$. For visualization purposes, the data is rescaled by the amplitude of the fit (maximum by a factor 1.12).

\subsection{GST Goodness-of-fit}
\label{sec:gst_fit}
An important aspect of the gate-set estimate provided by GST is the goodness-of-fit of the model's predictions to the measurement outcomes, that can be expressed in a number $N_\sigma$ which describes how many standard deviations the log-likelihood of the GST model lies above the statistically expected log-likelihood \cite{blume-kohout_demonstration_2017}. A system where each gate, and the SPAM, are described by stationary CPTP maps can be perfectly described by the GST model, and would thus result in $N_\sigma\approx1$ due to the inherent measurement uncertainty. Note that a good GST fit does not equate high gate fidelities. Experimental quantum systems rarely behave perfectly Markovian however, and generally, models with $N_\sigma\gg1$ are still considered informative~\cite{blume-kohout_demonstration_2017, madzik_precision_2022, bartling_universal_2025}. 

We observe that the model violation increases significantly for long sequences on all qubits. For the electron spin, despite measuring sequences up to $L=128$, we consider the data up to $L=16$ to provide a more conservative gate fidelity estimate, with $N_\sigma=15.6$ (see supp. Fig. \ref{fig:model_violation} for the data and model predictions for both $L$). The increased model violation may be attributed to a few causes. First, the observed decoherence under decoupling, of the form $\exp (-(N/N_0)^2)$ cannot be fully represented by a stationary CPTP map. Second, sample heating due to the applied control pulses, and in general any slow experimental drift on the timescale of the GST measurements ($\sim$30 minutes), such as fluctuating pulse amplitudes and magnetic field drift, can also result in model violation. For the nuclear spins we considered data up to $L=8$ ($L=2$) for $c_0$ ($c_1$), yielding model violations of $N_\sigma=31$. 

\subsection{Qubit-readout correction}
\label{sec:readout_cor}
All presented data is corrected for the characterized single-shot readout (SSRO) fidelity of the electron spin~\cite{robledo_high-fidelity_2011}. Prior to each measurement shot the charge state and frequency resonance of the NV centre is confirmed with a charge-resonance (CR) check~\cite{bernien_heralded_2013}. The nuclear-spin states are initialized with the room-temperature equipment by initializing the electron spin and swapping the state to the nuclear spin, making use of DDrf gates~\cite{taminiau_universal_2014, bradley_ten-qubit_2019}. Readout is performed similarly by mapping the $Z$ component of the nuclear-spin state to the electron-spin state~\cite{taminiau_universal_2014}.

To estimate the Bell-state fidelity (Fig. \ref{fig:bell}e), the two-qubit expectation values are also corrected for the added readout infidelity of the nuclear spin~\cite{bradley_ten-qubit_2019}. We run the same measurement sequence (Fig. \ref{fig:bell}d), but the CMOS plays no pulses and $c_1$ is readout in the $Z$-basis. Assuming the initialization and readout errors are equal, an approximation motivated by the fact that the same gate sequences are used for both, the measured $\left<Z\right>$ can be estimated by \begin{equation}
    \left<Z\right> = S^2_1,
\end{equation}
where $S_1$ is the initialization and readout error of $c_1$. The parities of the Bell-state can be corrected by dividing them by $S_1$ if the two-qubit parity has support on $c_1$ (i.e. no $I$ on $c_1$). We measure $\left<Z\right>=0.75$ so $S_1=0.866$.

\subsection{Estimation of two-qubit gate fidelity}
\label{sec:2q_fid}
To estimate the average gate fidelity of the two-qubit gate, we assume that the error can be described by a two-qubit depolarizing channel with error probability $p$ and assume all other gates are ideal. The initialization and readout infidelity derived in the previous section are included as single-qubit depolarizing channels on $c_1$. We find $p$ using a least-squares fit on the values of the three Pauli correlators that are expected to be non-zero. This yields the simulated data in Fig. \ref{fig:bell}e with $p=0.07(2)$. The gate fidelity follows from $F_{avg}=1-3p/4 = 0.95(2)$

\newcounter{chapter}
\renewcommand{\thechapter}{\Alph{chapter}}
\setcounter{chapter}{19}
\renewcommand{\thesection}{Supplementary Note \arabic{section}}
\setcounter{section}{0}

\renewcommand{\theequation}{S\arabic{section}.\arabic{equation}}
\setcounter{equation}{0}

\renewcommand{\thefigure}{S\arabic{figure}}
\setcounter{figure}{0}

\renewcommand{\thetable}{S\arabic{table}}
\setcounter{table}{0}
\counterwithin{figure}{chapter}

\bibliography{text/references_cmos, references}%
\clearpage
\onecolumngrid
\begin{center}
\textbf{\large Supplementary Figures}
\end{center}

\begin{figure}[h]
    \centering
    \includegraphics[width=\linewidth]{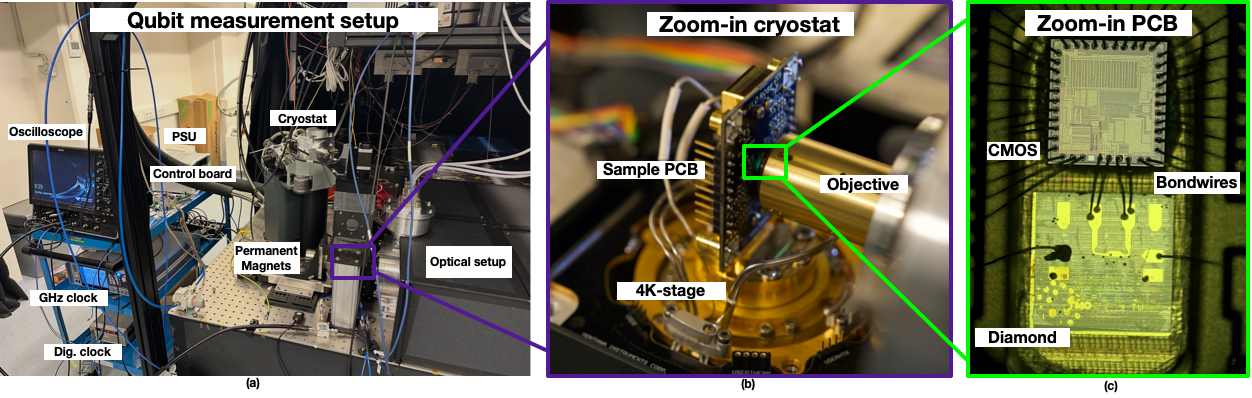}
    \caption{\emph{Pictures of the qubit control setup.} a) Qubit measurement setup with optical table and cryostat. b) Zoom-in of the cryostat before final assembly, with the PCB screwed to the cold-head but cables unconnected. c) Zoom-in of the chip assembly showing the cryo-CMOS control chip bonded to the diamond sample.}
    \label{fig:cmos_setup_pictures}
\end{figure}

\begin{figure}[h]
    \centering
    \includegraphics{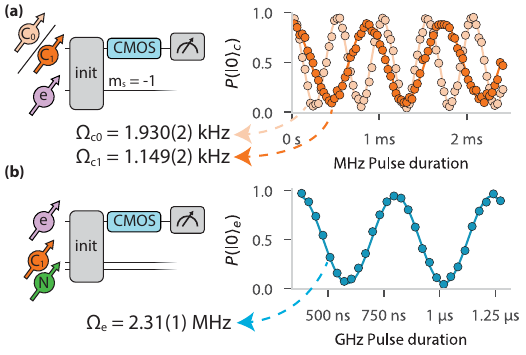}
    \caption{\emph{Rabi frequencies.} Data reproduced from our chip-characterization work~\cite{enthovenCryoCMOSElectronNuclearSpin2026}. \textbf{a)} Nuclear spin Rabi oscillations were measured with the electron in $m_s=-1$. \textbf{b)} The nitrogen spin and strongly-coupled carbon spin $c_0$ are initialized to enable full-contrast Rabi oscillations with the available Rabi frequency.}
    \label{fig:c_spin}
\end{figure}

\begin{figure}[h]
    \centering
    \includegraphics{figures/cmos_programme.png}
    \caption{\textbf{Example CMOS programme.} Visualisation of the CMOS sequencer programme that executes the final $L=128$ GST sequence for the electron spin. Certain instructions take more address space than others. The GHz signal generator is configured at the start of the experiment (before [a]). The controller then waits for a trigger, executes the programme, and at the end jumps back to the \textit{wait\_for\_trigger} instruction at [a]. Each logical gate consists of a $\pi/2$ gate and an XY8 decoupling sequence. The \textit{wr\_mw\_reg} instruction changes the selected pulse envelope and is used to switch between $\pi$ and $\pi/2$ pulses. Waiting times are performed here by repeated \textit{nop} instructions. Blocks of logical gates are repeated by using the auxiliary counter, which is reset at the start, and incremented at the end of a gate block. The \textit{jmp\_cond} instruction sets the instruction counter back to the start address of the block if the auxiliary counter value is smaller than the set number of repetitions.}
    \label{fig:cmos_programme}
\end{figure}

\begin{figure}[h]
    \centering
    \includegraphics{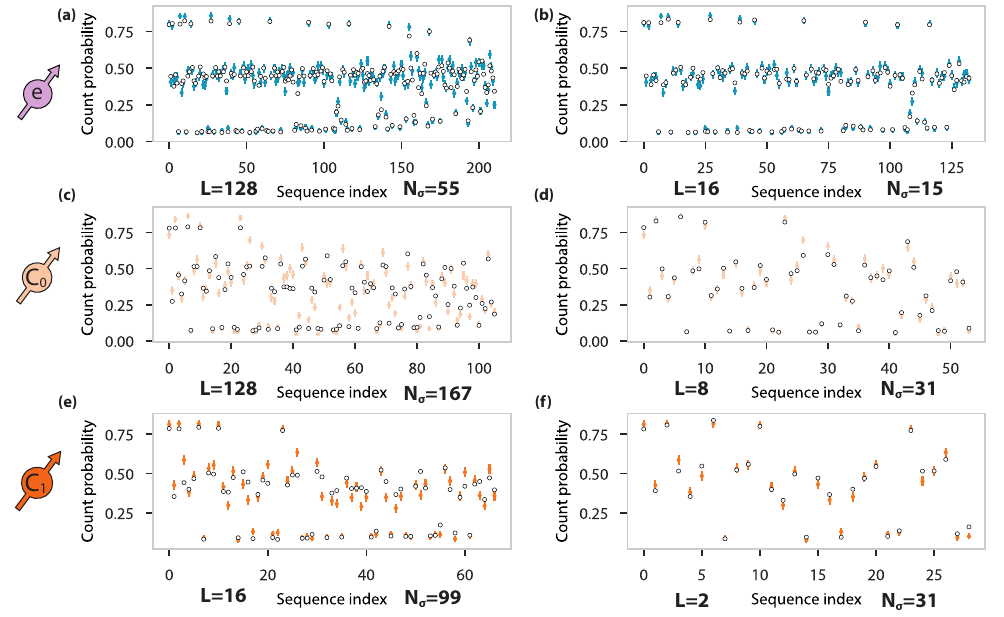}
    \caption{\textbf{GST model violation.} Solid, coloured circles are measurement data with measurement uncertainty errorbars. Open circles are the outcome predictions from the GST model. Left: Full dataset. Right: Truncated dataset used for reporting in this work. The models fit to higher $L$ data yielded higher gate fidelities, however their much higher model violation $N_\sigma$ means those fidelities are less trustworthy, and therefore we considered the shorter $L$ data instead.}
    \label{fig:model_violation}
\end{figure}

\begin{figure}[h]
    \centering
    \includegraphics{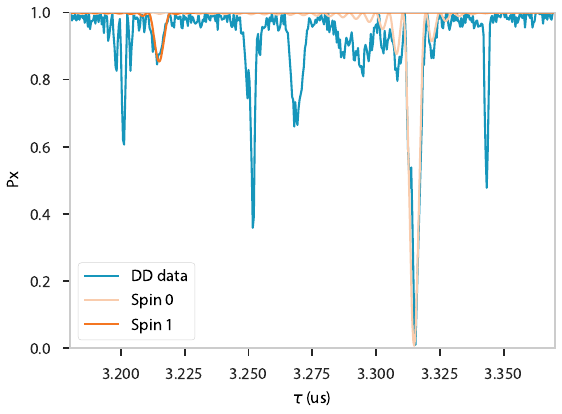}
    \caption{\textbf{DD data of nuclear spins.} Blue data: DD spectroscopy taken with room-temperature equipment, using $N=48$ decoupling $\pi$ pulses. The sequence is identical to the one shown in Fig. \ref{fig:e_spin}a. The dips in the $\tau$ spectrum indicate that the electron-spin interacts with one or more carbon spins. For the spins used in this work, the theoretical response was calculated~\cite{taminiau_detection_2012} and matched to the measured data to estimate $A_\parallel$ and $A_\perp$.}
    \label{fig:dd_extended}
\end{figure} %

\begin{figure}[h]
    \centering
    \includegraphics{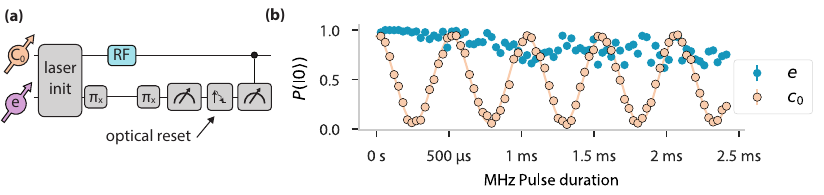}
    \caption{\textbf{Electron-spin population during nuclear-spin Rabi.} \textbf{a)} Sequence used for measuring the $c_0$ Rabi oscillations. Grey boxes are performed by RT equipment. Identical sequence as used for Fig. \ref{fig:c_spin}a. The electron spin is read out first, then reset, before the carbon spin is read out. \textbf{b)} Rabi data of $c_0$ (same as Fig. \ref{fig:c_spin}), now plotted together with electron-spin data. Ideal operation would result in the electron spin returning to the $\ket 0$ state. Likely due to heat generation by the CMOS chip, either the second $\pi$ pulse (that should return the electron spin from $\ket 1$ to $\ket 0$) is off-resonant, or the NV centre emits fewer photons under optical excitation. This could be caused by heat changing the magnetic field that is generated by the permanent magnet in close proximity to the sample or the NV centre optical transitions broadening at temperatures above 10~K\cite{fu_observation_2009}. The fact that the $c_0$ signal amplitude decays far less indicates that the electron-spin state is preserved during rf (MHz) driving (otherwise the MHz radiation would become off-resonant), that the electron-spin microwave pulses are resonant after the optical reset (as $c_0$ readout relies on electron-spin gates), and that the electron-spin readout is normal during the subsequent $c_0$ readout. Thus, the heating effect is highly transient and disappears in the time span from electron-spin readout and $c_0$ readout (typically $\sim$$1$ ms).}
    \label{fig:e_decay_rabi}
\end{figure}

\end{document}